\documentclass[twocolumn, unsortedaddress, superscriptaddress]{revtex4}
\usepackage{graphicx}
\usepackage{color}
\usepackage{amsmath}

\newcommand{\be}{\begin{equation}}
\newcommand{\ee}{\end{equation}}
\newcommand{\bea}{\begin{eqnarray}}
\newcommand{\eea}{\end{eqnarray}}

\begin{document}

\author{S.A. Jensen}
\affiliation{Max Planck Institute for
Polymer Research, Ackermannweg 10, 55128 Mainz, Germany}
\affiliation{FOM Institute AMOLF, Amsterdam, Science Park 104, 1098 XG Amsterdam, Netherlands}
\author{Z. Mics}
\affiliation{Max Planck Institute for
Polymer Research, Ackermannweg 10, 55128 Mainz, Germany}
\author{I. Ivanov}
\affiliation{Max Planck Institute for
Polymer Research, Ackermannweg 10, 55128 Mainz, Germany}
\author{H.S. Varol}
\affiliation{Max Planck Institute for
Polymer Research, Ackermannweg 10, 55128 Mainz, Germany}
\author{D. Turchinovich}
\affiliation{Max Planck Institute for
Polymer Research, Ackermannweg 10, 55128 Mainz, Germany}
\author{F.H.L. Koppens}\email{klaas-jan.tielrooij@icfo.es,   bonn@mpip-mainz.mpg.de, frank.koppens@icfo.es}
\affiliation{ICFO - Institut de
Ci\'encies Fot\'oniques, Mediterranean Technology Park,
Castelldefels (Barcelona) 08860, Spain}
\author{M. Bonn}\email{klaas-jan.tielrooij@icfo.es,   bonn@mpip-mainz.mpg.de, frank.koppens@icfo.es}
\affiliation{Max Planck Institute for
Polymer Research, Ackermannweg 10, 55128 Mainz, Germany}
\author{K.J. Tielrooij} \email{klaas-jan.tielrooij@icfo.es,   bonn@mpip-mainz.mpg.de, frank.koppens@icfo.es}
\affiliation{ICFO - Institut de
Ci\'encies Fot\'oniques, Mediterranean Technology Park,
Castelldefels (Barcelona) 08860, Spain}

\title{Competing Ultrafast Energy Relaxation Pathways in Photoexcited Graphene}

\newpage

\begin{abstract}
For most optoelectronic applications of graphene a thorough
understanding of the processes that govern energy relaxation of
photoexcited carriers is essential. The ultrafast energy relaxation in graphene occurs through two competing pathways: carrier-carrier scattering -- creating an elevated carrier temperature -- and optical phonon emission. At present, it is not clear what determines the dominating relaxation pathway. Here we reach a unifying picture of the ultrafast energy relaxation by investigating the terahertz photoconductivity, while varying the Fermi energy, photon energy, and fluence over a wide range. We find that sufficiently low fluence ($\lesssim$ 4 $\mu$J/cm$^2$) in conjunction with sufficiently high Fermi energy ($\gtrsim$ 0.1 eV) gives rise to energy relaxation that is dominated by carrier-carrier scattering, which leads to efficient
carrier heating. Upon increasing the fluence or decreasing the
Fermi energy, the carrier heating efficiency decreases, presumably due to energy relaxation that becomes increasingly dominated by
phonon emission. Carrier heating through carrier-carrier scattering
accounts for the negative photoconductivity for doped
graphene observed at terahertz frequencies. We present a simple
model that reproduces the data for a wide range of Fermi levels
and excitation energies, and allows us to qualitatively assess how
the branching ratio between the two distinct relaxation pathways
depends on excitation fluence and Fermi energy.
\\

\end{abstract}


\maketitle

Graphene is a promising material for, amongst others,
photo-sensing and  photovoltaic applications
\cite{Bonaccorso2010}, owing to its broadband absorption
\cite{Dawlaty2008, Ren2012}, its high carrier mobility
\cite{Novoselov2004, Wang2013} and the ability to create a photovoltage from
heated electrons or holes \cite{Gabor2011}. It furthermore
uniquely allows for electrical control of the carrier density and polarity \cite{Novoselov2004}. To
establish the potential and limitations of graphene-based
optoelectronic devices, a thorough understanding of the ultrafast
(sub-picosecond) primary energy relaxation dynamics of
photoexcited carriers is essential. For undoped graphene (with Fermi energy $E_F \approx$ 0), ultrafast energy relaxation through interband carrier-carrier scattering was predicted \cite{Winzer2010} and observed \cite{Brida2013, Ploetzing2014} to lead to multiple electron-hole pair excitation. For doped graphene (with Fermi energy $\lvert E_F \rvert >$ 0), ultrafast energy relaxation through carrier-carrier interaction also plays an important role, with intraband scattering leading to carrier heating \cite{Johannsen2013,Tielrooij2013,SongPRB2013,Gierz2013}. In addition, the ultrafast energy relaxation was ascribed to optical phonon emission \cite{Breusing2011, Lui2010}, which reduces the carrier heating efficiency. Closely related, for undoped graphene the sign of the terahertz (THz) photoconductivity is positive (see e.g.\ \cite{George2008, Strait2011, Shi2014, Frenzel2014}), whereas for
intrinsically doped graphene the sign is negative, meaning that
photoexcitation gives rise to an apparent decrease of conductivity
\cite{Docherty2012, Frenzel2013, Jnawali2013,
Tielrooij2013,Shi2014,Frenzel2014}. This negative photoconductivity
was attributed to stimulated THz emission \cite{Docherty2012}, and to a reduction of the intrinsic conductivity by enhanced scattering with optical phonons \cite{Jnawali2013, Frenzel2013} or by carrier heating \cite{Tielrooij2013, Shi2014, Frenzel2014}. Since the experimental parameters of all these studies differ strongly, different conclusions were drawn regarding the dominating energy relaxation pathway and the origin of the sign of the THz photoconductivity.
\\

Here, we reach a unifying picture of the energy relaxation of photoexcited carriers in graphene by experimentally studying the ultrafast energy relaxation of photoexcited carriers for a large parameter space, where we vary the Fermi energy and the fluence. We compare the data with a simple model of carrier heating, which quantitatively reproduces the frequency-dependent THz photoconductivity for a large range of Fermi energies and fluences with a single free fit parameter. This fit parameter is the carrier heating efficiency, i.e.\ the fraction of absorbed energy from
incident light that is transferred to the electron system.
Although the THz photoconductivity is not a direct probe of the
carrier temperature, and the model parameters carry some
uncertainty, this approach allows us to identify qualitatively how
the ultrafast energy relaxation processes depend on excitation
fluence and Fermi energy.
\\

We study the ultrafast carrier-energy relaxation using
time-resolved  THz spectroscopy experiments, where an excitation
pulse with a wavelength of 400, 800 or 1500 nm (corresponding to a
photon energy of $E_0$ = 3.10, 1.55 or 0.83 eV, respectively)
excites electron-hole pairs, and a low-energy terahertz (THz, $f=
0.3-2$ THz) probe pulse interrogates the sample, where it
interacts with mobile carriers. An optical delay line controls the
relative time delay between excitation and probe pulses, 
\\
\\

making it
possible to determine the photo-induced change in conductivity
(the photoconductivity $\Delta \sigma$) as a function of time with
a time resolution of $\sim$150 fs \cite{Ulbricht2011}. We use two
different samples: The first sample, where the graphene has a controllable Fermi energy, contains Chemical Vapor Deposition
(CVD) grown graphene with an area of 1 cm$^2$ transferred onto a substrate that consists of
doped silicon, covered by a 300 nm thick layer of SiO$_2$. The silicon
serves as the backgate for electrical control of the Fermi energy
(carrier doping) of the graphene sheet from $E_F = 0.3$ eV down to
$\sim$0.06 eV (the width of the neutrality region is $\sim$0.06
eV, see Supp.\ Info). We use weakly doped silicon (resistivity of
10-20 $\Omega$cm) and sufficiently low pump energy (0.83 eV) to
ensure that the photo-induced signal from the silicon in the
substrate (without graphene) is minimized (See Supp.\ Info). Two
silver-pasted electrical contacts to the graphene sheet enable
resistance measurements to retrieve the Fermi energy through the
capacitive coupling of the backgate (See Supp.\ Info). The second sample
contains CVD grown graphene with an area of a square inch transferred onto SiO$_2$ with a fixed
Fermi level of $<$0.15 eV (see Supp.\ Info). We carefully measure
the fluence $F$ as described in Ref.\ \cite{Pijpers2009} and use
the experimentally determined optical absorption of graphene on
SiO$_2$ using standard UV-Vis spectrometry to obtain the absorbed
fluence $F_{\rm abs}$, and from this the number of absorbed
photons $N_{\rm exc} = F_{\rm abs}/E_0$ , which is equal to the number of primary excited carriers.
\\

To investigate the ultrafast (sub-picosecond) energy relaxation
dynamics of photoexcited carriers, we first examine the temporal
evolution of the photoconductivity. In Fig.\ 1a we show the
dynamics for a range of Fermi energies using the sample with controllable Fermi energy, and in Fig.\ 1b for a range of
fluences using the sample with fixed Fermi energy. We find that
all traces, except the one with Fermi energy within the neutrality
region width ($E_F < 0.06$ eV), show negative photoconductivity
with a sub-picosecond rise, followed by a picosecond decay.
Similar pump-probe dynamics have been observed before
\cite{Tielrooij2013,Strait2011, Jnawali2013, George2008,
Frenzel2013, Frenzel2014, Docherty2012, Shi2014} and can be
understood as follows: During the rise of the signal three main
processes take place: \textit{(i)} the creation of initial
electron-hole pairs; and subsequent ultrafast energy relaxation
through two competing relaxation channels, namely \textit{(ii)}
carrier-carrier scattering and \textit{(iii)} optical phonon
emission. Thus, the peak signal corresponds to a 'hot state' with
an elevated carrier temperature $T_{\rm el}$ and/or more energy in
optical phonons \cite{Brida2013,SongPRB2013, Johannsen2013,
Gierz2013, Lui2010, Tielrooij2013, Breusing2011}. During the
subsequent picosecond decay, the 'hot state' cools down to the
same state as before photoexcitation. The fraction of absorbed
energy that -- after the initial ultrafast energy relaxation -- ends up in the electron system or in the phonon system
depends on the timescales associated with carrier heating and
phonon emission, respectively.
\\

The ultrafast energy relaxation takes place during the first few
hundred femtoseconds after photoexcitation, i.e. during the rise
of the conductivity change. Figs.\ 1c and 1d show the normalized
photoconductivity signals for this time window. We will discuss
the evolution of the signal amplitude as a function of Fermi
energy and fluence later. First we note that the rise dynamics
exhibit an intriguing effect: upon decreasing the Fermi energy
(i.e.\ the density of intrinsic carriers $N_{\rm int}$) or
increasing the fluence (i.e.\ the density of primary excited
carriers $N_{\rm exc}$) the signal peak is reached at increasingly
later times. To quantify these results, we describe the dynamics using two rise times and an exponential decay time. The two rise times allow for part of the conductivity change to occur with the (fixed) experimental time resolution and part with a (free) slower time scale. We then examine the effective rise time $\tau_{\rm rise}$, which is the amplitude-weighted average of the two (see Methods). The
insets of Fig.\ 1c and Fig.\ 1d show $\tau_{\rm rise}$ as a
function of $N_{\rm int}$ and $N_{\rm exc}$, respectively. Indeed,
for decreasing $N_{\rm int}$ and increasing $N_{\rm exc}$ the
effective rise time increases from below 200 fs (limited by the
experimental time resolution) up to 400 fs (for $E_F <$0.1 eV).
\\

The slowing down of the ultrafast energy relaxation of
photoexcited carriers with decreasing Fermi energy is consistent with energy relaxation through intraband carrier-carrier scattering (see bottom right inset in Fig.\ 1c). The microscopic picture of this scattering process is shown for two different Fermi energies in the top left inset of Fig.\ 1c. Photoexcited carriers relax by exchanging energy with intrinsic conduction band carriers that thus heat up. The amount of energy that is exchanged between photoexcited carriers and intrinsic conduction band carriers in each intraband carrier-carrier scattering event is $\sim E_F$ \cite{SongPRB2013}. Therefore, if $E_F$ decreases, more energy-exchange events are
required for the photoexcited carriers to complete their energy
relaxation cascade and therefore the relaxation time will
increase (see bottom right inset of Fig.\ 1c). If the energy relaxation through carrier-carrier scattering would slow down in such a way that the relaxation rate becomes comparable to the rate of other relaxation channels
(e.g.\ optical phonon emission), we expect these channels to start
contributing to the overall energy relaxation. This would lead to
a decrease in the fraction of energy that is transferred to the
electron system, i.e.\ a reduced carrier heating efficiency.
\\

We quantify the fraction of absorbed energy that leads to carrier heating by comparing
our THz photoconductivity data with the results of a simple
thermodynamic model. In short,
carrier heating leads to a broader carrier distribution (higher $T_{\rm el}$), which -- in combination with an energy-dependent scattering time
\cite{ando,dassarmareview} -- leads to the photo-induced change in THz conductivity (see Methods). The carrier heating is governed by the amount of absorbed energy $F_{\rm abs}$ and the electronic heat capacity of graphene, which for a degenerate electron gas is given by $C_{\rm el} = \alpha T_{\rm el}$ \cite{Kittel}. Here $\alpha = {{2 \pi E_F}\over{3 \hbar^2 v_F^2}} k_B^2$, with $\hbar$, $v_F$ and $k_B$ the reduced Planck constant, the Fermi velocity and Boltzmann's constant, respectively. The possibility of controlling the Fermi energy of graphene thus allows for tunability of the heat capacity (see Fig.\ 2a), which in turn determines the 'hot' carrier temperature $T'_{\rm el}$ that the system reaches. Figure 2b shows that $T'_{\rm el}$ is equivalently determined by both the fluence and the Fermi energy.
\\

To calculate the 'hot' carrier temperature we use a basic numerical approach, which produces a heat capacity in the degenerate regime that corresponds well with the analytical heat capacity (see Fig.\ 2a), while remaining valid for non-degenerate electron temperatures ($k_B T_{\rm el} > E_F$). The numerical approach is based on the concept that before photoexcitation there is a known amount of energy in the electronic system: $\mathcal{E}_1 = \int_0^\infty d\epsilon D(\epsilon) \epsilon F(E_F, T_{\rm el})$; and a known number of
carriers in the conduction band: $N_{\rm int} = \int_0^\infty
d\epsilon D(\epsilon)F(E_F, T_{\rm el})$, where $D(\epsilon)$ is
the energy-dependent density of states and $F(E_F, T_{\rm el})$ is
the Fermi-Dirac distribution that depends on Fermi energy $E_F$
and carrier temperature $T_{\rm el}$. Due to optical excitation,
an amount of energy $F_{\rm abs}$ is absorbed in the graphene and
a fraction $\eta$ of this energy ends up in the electronic system
trough intraband carrier-carrier scattering. After intraband
heating is complete, the system is then described by the following set of
equations: $\mathcal{E}_2 = \mathcal{E}_1 + \eta F_{\rm abs} = \int_0^\infty
d\epsilon D(\epsilon) \epsilon F(E'_F, T'_{\rm el})$, and
(conserving the number of conduction band carriers) $N_{\rm int} =
\int_0^\infty d\epsilon D(\epsilon)F(E'_F, T'_{\rm el})$. Here
$E'_F$ and $T'_{\rm el}$ are the chemical potential and the
carrier temperature in the 'hot state', respectively. Carrier
heating thus alters the carrier distribution, where we find by numerically solving the equations for $\mathcal{E}_2$ and $N_{\rm int}$ that the carrier
temperature increases and the chemical potential decreases by photoexcitation (see inset of Fig.\ 2c and Methods). The photo-induced increase of carrier temperature, and the associated decrease in chemical potential were experimentally confirmed recently \cite{Gierz2013}.
\\

The hot carrier distribution, with $E'_F$ and $T'_{\rm el}$ calculated using the carrier heating model, directly leads to negative THz photoconductivity (see Methods), which scales linearly with carrier temperature up to $\sim$2000 K and then shows some saturation behavior (Fig.\ 2c). We test the validity of our carrier heating model by comparing its predictions for the frequency-resolved
photoconductivity with our experimental results for the sample with fixed Fermi energy. In
Fig.\ 3a we show this comparison for a fluence of $\sim$12
$\mu$J/cm$^2$ (pump wavelength 800 nm, $N_{\rm exc} =
1\cdot10^{12}$ absorbed photons/cm$^2$). We compare the data (at the time delay that corresponds to the signal peak) with
the model result for a ground state Fermi energy of 0.11 eV and a scattering
time proportionality constant of 200 fs/eV (extracted from the Raman spectrum, THz conductivity measurements on the same sample without photoexcitation, and the saturation value of the THz photoconductivity at high fluence; see also Supp.\ Info) and find good agreement with a heating efficiency of $\eta$ = 0.75. The small discrepancies between data and model can be ascribed to artifacts that arise from the temporal change of the photoconductivity during the interaction with the THz pulse \cite{Hendry2005}, although we largely avoid these by moving the optical pump delay line simultaneously with the THz probe delay line.
\\

The overall agreement between data and model shows that the observed negative
THz photoconductivity of intrinsically doped graphene
\cite{Docherty2012, Frenzel2013, Jnawali2013, Tielrooij2013,
Shi2014, Frenzel2014} can be fully reproduced by considering intraband carrier
heating, which reduces the thermally averaged conductivity of the
intrinsic carriers. Despite the simplicity of the model, it can
also explain the experimental results in Ref.\ \cite{Frenzel2013}
using their experimental parameters, as well as the results in
Ref.\ \cite{Docherty2012}, by letting the environmental gas change
the Fermi energy. These observations lead us to conclude that, despite some
uncertainty in the determination of the Fermi energy and the
scattering time, the model is suitable for obtaining reliable
qualitative indications on how the carrier heating efficiency
depends on the Fermi energy and the fluence.
\\

We now examine the 'high fluence' regime using  $N_{\rm exc} =
8\cdot10^{12}$ absorbed photons/cm$^2$ (a fluence of $\sim$100
$\mu$J/cm$^2$) in Fig.\ 3b. This corresponds to the regime where
the energy relaxation is slower than the experimental time
resolution (see Fig.\ 1d). Here we find that we can only describe
the data with a significantly reduced carrier heating efficiency
of $\eta \approx$ 0.5 (keeping the ground state Fermi energy and scattering
time proportionality constant the same as in the low fluence regime). Combined, these
results show that at sufficiently low fluence, a large fraction of
the absorbed energy ends up in the electron system, i.e.\ the
ultrafast energy relaxation occurs through efficient (and fast)
carrier-carrier scattering. However, upon increasing the fluence
(i.e.\ the carrier temperature), the relative amount of energy
transferred to the electron system decreases, which means that
carrier-carrier scattering becomes less efficient (and slower) and
other relaxation processes start to
contribute.
\\

To determine in more detail how the carrier heating efficiency
depends on fluence, we study the peak photoconductivity of the sample with fixed Fermi energy for a large range of excitation powers, for both 800 nm and 400
nm excitation. We show the photoconductivity at the peak (when the
ultrafast energy relaxation is complete) in Fig.\ 4a-b, together
with the results of the carrier heating model for the same
parameters as in Fig.\ 3a and a frequency of 0.7 THz. For low fluences (up to $\sim$4
$\mu$J/cm$^2$, corresponding to $N_{\rm exc} \sim 0.3\cdot10^{12}$
absorbed photons/cm$^2$, 800 nm excitation, $\sim$2\% absorption)
the experimental data are in agreement with the heating model with
a fixed heating efficiency of $\eta =$ 1. We notice that the
calculated photoconductivity shows saturation behavior with
$N_{\rm exc}$ even for constant $\eta$. This is related to the
nonlinear dependence of the THz photoconductivity on carrier
temperature (Fig.\ 2c). Interestingly, at fluences above $\sim$4
$\mu$J/cm$^2$ the experimental photoconductivity starts saturating
and the model is only in agreement for a heating efficiency that
gradually decreases to $\sim$50\% for the highest fluences applied
here (Fig.\ 4a). These observations suggest that once a certain
carrier temperature ($\sim$4000 K, see inset Fig.\ 4b) is reached,
the heating efficiency decreases. Interestingly, the experimental
data for excitation with 400 nm light start deviating from the
model (with efficient heating) at $N_{\rm exc}
\sim$0.15$\cdot10^{12}$ absorbed photons/cm$^2$ (Fig.\ 4b),
instead of $\sim$0.3$\cdot10^{12}$ absorbed photons/cm$^2$ in the
case of excitation with 800 nm light. This is because each 400 nm
photon has twice the energy of a 800 nm photon. Thus, in both
cases the carrier heating efficiency starts decreasing around the
same carrier temperature.
\\

We now determine how the heating efficiency depends  on the Fermi
energy by measuring the peak photoconductivity for the sample with controllable Fermi energy as a
function of both excitation power and Fermi energy. The combined
results (for excitation at 1500 nm) are represented in Fig.\ 5a,
where we show the peak photoconductivity as a function of gate
voltage for five different excitation powers. In Fig.\ 5b we show
the peak photoconductivity as a function of $N_{\rm exc}$ for
three distinct gate voltages, corresponding to $N_{\rm int}
\approx$ 1, 2 and 3$\cdot10^{12}$ carriers/cm$^2$, and compare
these to the results of the heating model. We find that for a
doping of $N_{\rm int}$ = 3$\cdot10^{12}$ carriers/cm$^2$ ($E_F
\sim$0.2 eV) the data is in good agreement with the carrier
heating model, using a scattering time proportionality constant of $\sim$50 fs/eV and a
carrier heating efficiency of $\eta$ =1. However, for the lowest
$N_{\rm int}$, which corresponds to a Fermi energy of $\sim$0.1
eV, we find a carrier heating efficiency of $\sim$20\%, using the
same energy-dependent scattering time. These results are in good
agreement with the observed slowdown of the rise dynamics with
decreasing Fermi energy in Fig.\ 1b.
\\

Comparing the data and the heating model leads to the following physical picture of the ultrafast energy relaxation in graphene: Until a certain carrier temperature is reached ($\sim$4000 K), the ultrafast energy relaxation is dominated by carrier-carrier scattering, which leads to efficient and fast ($<$150 fs) carrier heating. Once this carrier
temperature is reached, the relaxation slows down and the carrier
heating efficiency decreases, as ultrafast energy relaxation
occurs through additional pathways involving optical phonon
emission \cite{Lui2010}. The reduction in heating efficiency that follows from the macroscopic heating model can be explained using the microscopic picture of intraband carrier-carrier scattering, as put forward in Refs.\ \cite{Tielrooij2013, SongPRB2013}. At increased electron temperatures, the
(quasi-equilibrium) Fermi energy decreases (see Methods), which means that the electronic heat capacity decreases. It furthermore implies that the amount of energy that is exchanged in intraband carrier-carrier scattering events ($\sim E_F$) decreases. Therefore, energy relaxation of a photoexcited carrier requires an increasing number of intraband carier-carrier
scattering cascade steps. Thus for an increasing carrier temperature, energy relaxation through intraband carrier heating slows down.
\\

The physical picture of carrier-temperature dependent
ultrafast energy relaxation of photoexcited carriers in graphene
unites the conclusions of a large fraction of the existing
literature on this topic. For example in Ref.\ \cite{Lui2010},
with excitation in the 'high fluence' regime ($\sim$10$^{14}$
absorbed photons/cm$^2$ at 800 nm), it was concluded from the
experimentally measured carrier temperature that only part of the
absorbed light energy ends up in the electronic system, whereas
the rest couples to optical phonons. An ultrafast optical
pump-probe study employing a fluence of 200 $\mu$J/cm$^2$
\cite{Breusing2011}, also in the 'high fluence' regime, similarly
demonstrated optical phonon mediated relaxation, in addition to
carrier heating. By measuring in the 'low fluence' regime (a few $\mu$J's), Ref.\ \cite{Tielrooij2013} concluded that the ultrafast energy
relaxation was dominated by carrier heating. Furthermore, two very
recent optical pump - THz probe studies \cite{Shi2014,Frenzel2014}
both used an excited carrier density of $N_{\rm exc} \approx 1
\cdot 10^{12}$ absorbed photon/cm$^{2}$ at 800 nm and ascribe
their observed negative THz photoconductivity (partially) to
carrier heating.
\\

In conclusion, we provide a unifying explanation of  the ultrafast
energy relaxation of photoexcited carriers in graphene. For
sufficiently low excitation power and sufficiently high Fermi
energy, the relaxation is dominated by carrier-carrier scattering,
which leads to efficient generation of hot carriers. This regime typically persists up to a fluence of $\sim$4 $\mu$J/cm$^2$ (or $N_{\rm exc} = 0.3\cdot10^{12}$ absorbed 800 nm photons/cm$^2$) for a Fermi energy of 0.1 eV. For larger Fermi energy, a higher fluence can be used without significant reduction in heating efficiency. In the case of lower Fermi energy and/or a higher fluence, the heating efficiency will
decrease due to slower intraband carrier-carrier scattering and additional energy relaxation channels involving optical phonon emission. This opens up the
possibility to control the pathway of ultrafast energy relaxation,
i.e.\ the ability to tune the efficiency of energy transfer from
the primary excited carriers to electronic heat or to alternative
degrees of freedom, such as lattice heat. Such tunability is
useful for future applications, for instance in the field of
photodetection, where hot carriers are the dominant source of photocurrent generation \cite{Gabor2011}. Finally, we note that the terrestrial solar
radiation (on the order of a pJ/cm$^2$ during a 10 ps timescale)
corresponds to the 'low fluence' regime with efficient carrier
heating, which is therefore the relevant process to consider for
photovoltaic applications.

\section*{METHODS}

\subsection*{Photoconductivity from carrier heating model}

We use a numerical model based on carrier heating to calculate the
complex photoconductivity of photoexcited graphene. The frequency-dependent
conductivity of graphene is generally given by \cite{dassarmareview}

\be \sigma (\omega) = {{e^2 v_F^2}\over{2}} \int_0^\infty
d\epsilon D(\epsilon) {{\tau_{\rm
scatter}(\epsilon)}\over{1-i\omega\tau_{\rm scatter}(\epsilon)}}
{{dF(E_F, T_{\rm el})}\over{d\epsilon}} ,\ee

with $e$ the elementary charge, $v_F$ the Fermi velocity,
$D(\epsilon) = {{2\epsilon}\over{\pi \hbar^2 v_F^2}}$ the density
of states, $\hbar$ the reduced Planck constant, $\tau_{\rm
scatter} (\epsilon)$ the energy-dependent momentum scattering time
and $F(E_F, T_{\rm el})$ the Fermi-Dirac distribution that is
determined by the Fermi level and the carrier temperature. For
unexcited graphene we use the steady state Fermi level $E_F$ and
the ambient temperature $T_{\rm el}$. We furthermore use a scattering time
that is determined by charged impurity scattering and increases linearly
with energy $\epsilon$ \cite{ando,dassarmareview}. After electron-hole pair
excitation of graphene, photoexcited carriers can interact with
the intrinsic carriers, leading to intraband thermalization (see also main text): the
carrier temperature increases to $T_{\rm el}'$ and the chemical
potential decreases to $E'_F$. The reason for the decrease of the chemical potential is the linear scaling of the $D (\epsilon)$ with energy: a broader carrier distribution (higher carrier temperature) would lead to an increased number of carriers in the conduction band, if the Fermi energy would be kept constant. Therefore, a higher carrier temperature leads to a lower Fermi energy, as confirmed experimentally in Ref.\ \cite{Gierz2013}. The conductivity of photoexcited graphene $\sigma' (\omega)$ then follows from Eq.\ 1, using the 'hot Fermi level' and 'hot carrier temperature' with the photoconductivity given by $\Delta \sigma (\omega) = \sigma'
(\omega) - \sigma (\omega)$.
\\

Our model takes into account the effect of intraband carrier heating on the photoconductivity through the carrier distribution, which is sufficient to explain the observed negative photoconductivity \cite{Docherty2012, Frenzel2013,Jnawali2013, Tielrooij2013,Shi2014,Frenzel2014} and the dependence on excitation power and Fermi energy. The model does not explicitly include the effect of energy relaxation to optical
phonons on the photoconductivity, as proposed in Ref.\
\cite{Jnawali2013}. This is justified, because for
the graphene used here (with an impurity-scattering-limited mobility below 2500 cm$^2$/Vs) the effect of phonons on the conductivity is negligibly small at low fluences \cite{Tielrooij2013}. We note that for very high fluences ($N_{\rm exc} > 10^{12}$ absorbed photons/cm$^2$) and for phonon-scattering-limited graphene with mobilities $>$ 10,000 cm$^2$/Vs \cite{Perebeinos2010}, this effect likely plays a role, in addition to carrier heating. Furthermore, a more advanced model could include deviations from linear scaling between the scattering time and the carrier energy \cite{Hwang2007}, as well as changes in the Drude weight. The latter effect occurs when the valence and conduction band electrons no longer have separate thermal distributions and likely plays an important role around the Dirac point, where it correctly produces positive photoconductivity  \cite{Frenzel2014}.

\subsection*{Rise dynamics}

We describe the time-resolved photoconductity with a
phenomenological  model that includes a rise step and an
exponential decay step with time $\tau_{\rm decay}$. In the low
excitation power/high Fermi energy regime (highly efficient
carrier heating), the rise occurs within our instrument response
function. For high excitation power/low Fermi energy the rise
contains a very fast component and a slow component (reduced
carrier heating efficiency). Therefore we model the rise dynamics
with a 'fast' rise component that is fixed at a value equal to the
pulse duration of $\tau_{\rm pulse}$ = 120--150 fs (depending on
the excitation wavelength) and a variable 'slow' rise component
$\tau_{\rm slow}$ with a longer pulse duration, and describe the
time-resolved photoconductivity by

\begin{eqnarray} -\Delta \sigma  (t) = A \cdot {\rm Conv}({\rm
e}^{-t/{\tau_{\rm decay}}}, \tau_{\rm pulse}) \hspace{1mm} + \nonumber \\
B \cdot {\rm
Conv}({\rm e}^{-t/{\tau_{\rm decay}}}, \tau_{\rm slow})  , \end{eqnarray}

where $A$ and $B$ are the amplitudes corresponding to the 'fast'
and the 'slow' rise components and Conv$(y(t), \tau)$ means taking
the convolution of $y(t)$ with a Gaussian pulse of width $\tau$.
The effective rise time is then given by

\be \tau_{\rm rise} = {{A + B}\over{A/\tau_{\rm pulse} +
B/\tau_{\rm slow}}}   .\ee

In the fit, the free parameters are $A$, $B$, $\tau_{\rm slow}$
and $\tau_{\rm decay}$. The resulting effective rise time gives an
indication of how fast the initial energy relaxation of
photoexcited carriers takes place.

\section*{ACKNOWLEDGEMENTS}

We would like  to thank Justin Song,
Leonid Levitov, Sebastien Nanot and Enrique C\'anovas for useful discussions. KJT thanks NWO for a Rubicon fellowship. FK acknowledges support by the Fundacio Cellex Barcelona, the ERC Career integration grant 294056 (GRANOP), the ERC starting grant 307806 (CarbonLight) and support by the E.\ C.\ under Graphene Flagship (contract no. CNECT-ICT-604391).
\\

\clearpage

\onecolumngrid

\section{FIGURES}

\begin{figure} [h!!!!!]
   \centering
   \includegraphics [scale=0.8]
   {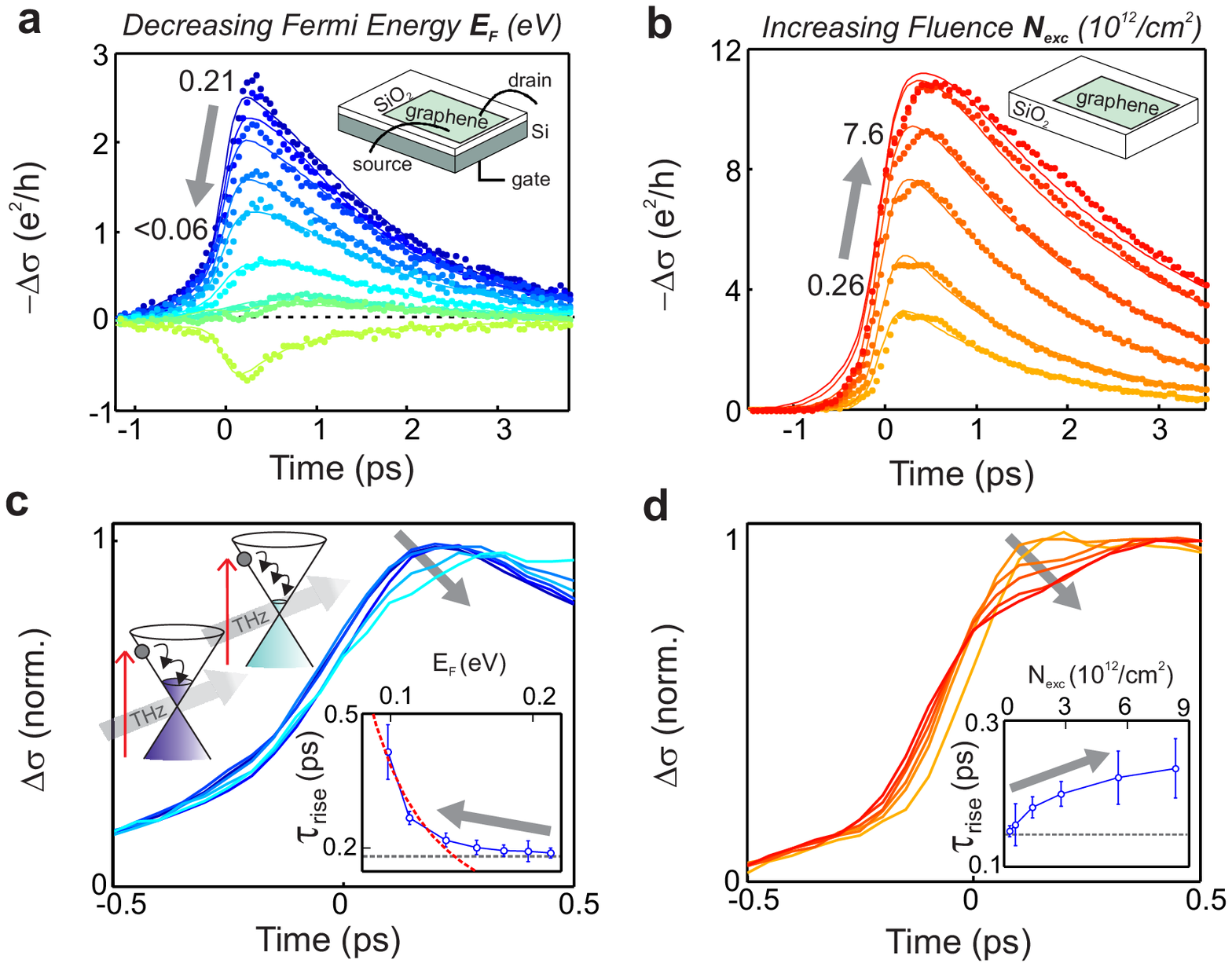}
   \caption{\textit{Energy relaxation dynamics - dependence on Fermi energy
and fluence.} \textbf{a)} The time-resolved photoconductivity for nine different gate voltages ($V - V_D$ =
-70, -60, -50, -40, -30, -20, -15, -10 and 0 V), i.e.\ for
decreasing Fermi energy $E_F$ or intrinsic carrier concentration
$N_{\rm int}$ (sample with controllable Fermi energy). The photoconductivity is extracted from
the measured change in THz transmission after photoexcitation with
1500 nm light (excitation power corresponds to $N_{\rm exc} =
0.8\cdot10^{12}$ absorbed photons/cm$^2$), using the thin film
approximation as in Ref.\ \cite{Tielrooij2013}. The inset shows
the geometry of the sample, as
explained in the text. The photoconductivity is negative for all
traces, except for the one closest to the Dirac point, where it is
positive. The solid lines are fits with an effective rise time
$\tau_{\rm rise}$ and an exponential decay time (see Methods).
\textbf{b)} The time-resolved photoconductivity for excitation with 800 nm light at six different
excitation powers, corresponding to absorbed photon densities of
$N_{\rm exc}$ = 0.26, 0.50, 1.2, 2.5, 5.0 and 7.6$\cdot10^{12}$
photons/cm$^2$ (sample with fixed Fermi energy). The geometry of the sample with a fixed
Fermi energy of $E_F \sim$0.1 eV (see Supp.\ Info), is shown in
the inset and explained in the text. The photoconductivity is
negative and increases for increasing excitation power, as the
carrier temperature increases. The solid lines are fits with an
effective rise time $\tau_{\rm rise}$ and an exponential decay
time (see Methods). \textbf{c)} The photoconductivity normalized
to the peak of the signal (sample with controllable Fermi energy), showing that decreasing the Fermi energy
leads to slower rise dynamics (i.e.\ slower energy relaxation).
The fitted values for $\tau_{\rm rise}$ as a function of $N_{\rm
int}$ are shown in the inset, together with the theoretically
predicted energy relaxation time based on Ref.\ \cite{SongPRB2013}
(scaled by a factor 1.8) and the experimental time resolution (horizontal dashed line). The schematics in the top left show the
ultrafast energy relaxation of a photoexcited carrier for two
different Fermi energies, according to the carrier heating process
described in Refs.\ \cite{Tielrooij2013, SongPRB2013}. \textbf{d)}
The photoconductivity normalized to the peak of the signal (sample with fixed Fermi energy), showing
that increasing the excitation power leads to slower rise
dynamics. The fitted values for $\tau_{\rm rise}$ as a function of
$N_{\rm exc}$ are shown in the inset together with the experimental time resolution (horizontal dashed line).}
   \end{figure}

\clearpage

\begin{figure} [h!!!!!]
   \centering
   \includegraphics [scale=1]
   {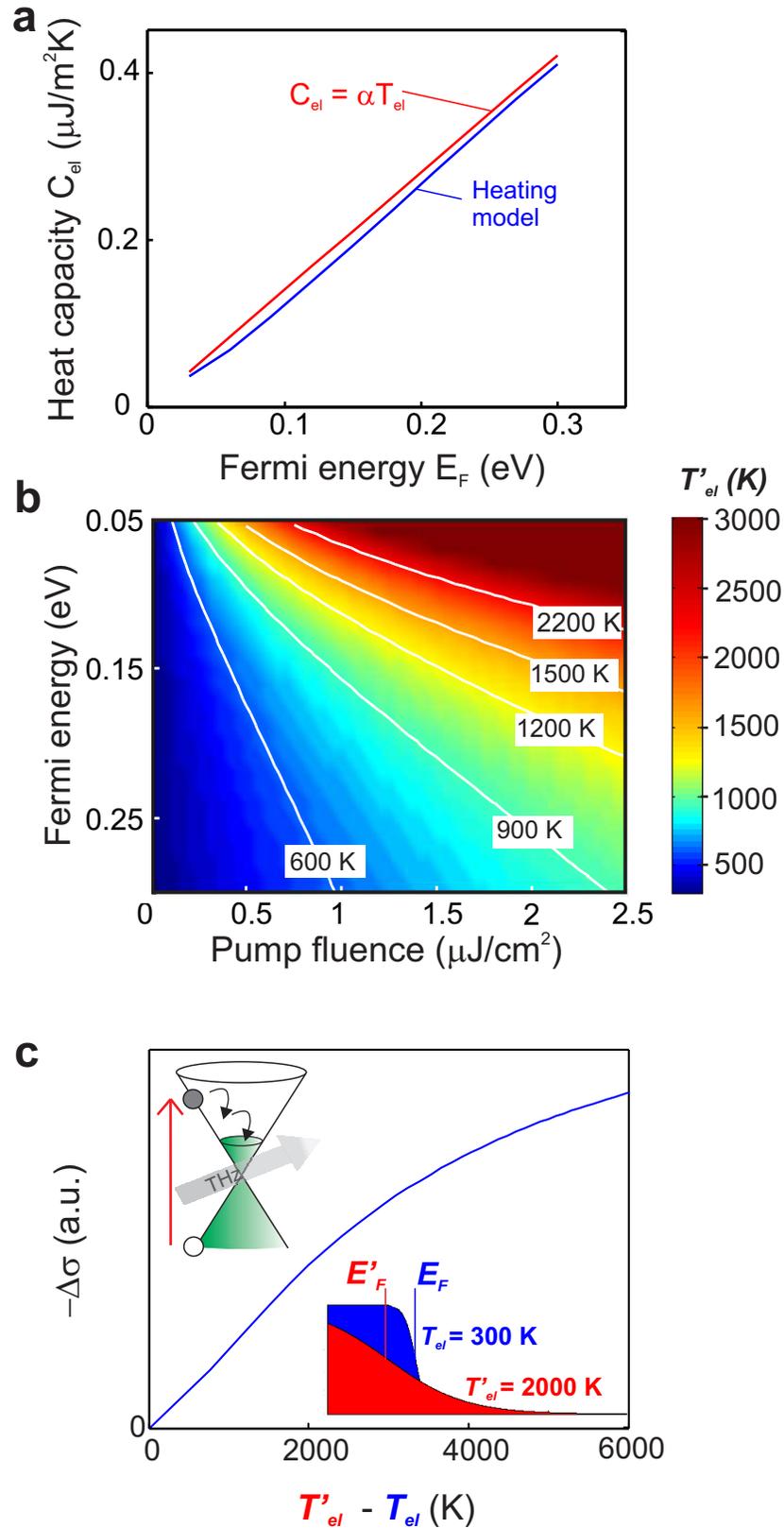}
      \caption{\textit{Carrier heating model.} \textbf{a)} The electronic heat capacity according to the analytical theory (red line) and our numerical model (blue line), showing linear scaling with Fermi energy. \textbf{b)}
Contour plot of the temperature of the carrier population,
showing that decreasing the Fermi energy and increasing the
fluence have an equivalent effect on the temperature of the
electrons after photoexcitation and intraband thermalization. The
carrier temperature is calculated through a simple thermodynamic
model (see text). \textbf{c)} The calculated negative THz photoconductivity
as a function of carrier temperature, showing a linear
dependence up to $\sim$2000 K. The top left inset shows schematically the process of
ultrafast energy relaxation of a photoexcited carrier, which leads
to an increase of the carrier temperature from $T_{\rm el}$ to
$T'_{\rm el}$. The inset on the bottom right shows the Fermi-Dirac
distribution of the conduction band carriers, where the
temperature increase leads to a decrease of the Fermi energy (to
keep the total number of conduction band carriers constant).}
\end{figure}

\clearpage

\begin{figure} [h!!!!!]
   \centering
   \includegraphics [scale=1.2]
   {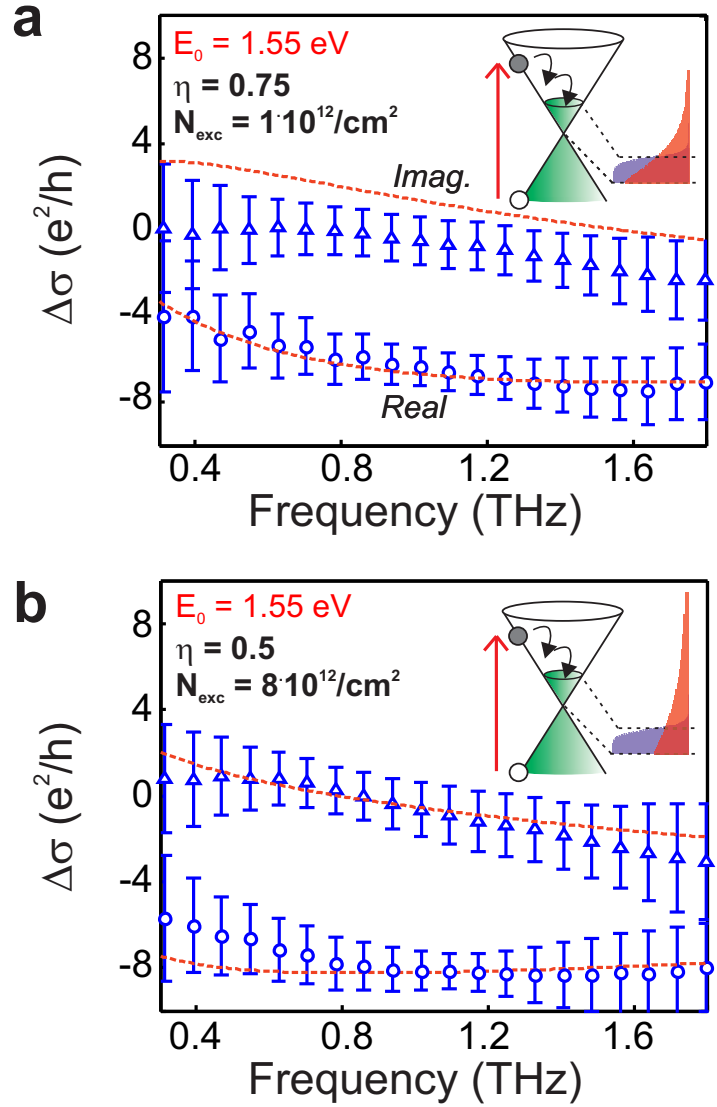}
      \caption{\textit{Comparison of experimental and theoretical photoconductivity.}
\textbf{a)} The complex photoconductivity of the sample with fixed Fermi energy as a function of frequency for an excitation fluence
corresponding to $N_{\rm exc} = 1\cdot10^{12}$ carriers/cm$^2$,
together with the model result with a carrier heating efficiency
of $\eta$ = 0.75. The model (see also main text and Methods)
describes the photoconductivity after photoexcited carriers have
thermalized through intraband carrier-carrier scattering and
directly yields the observed negative photoconductivity.
\textbf{b)} The complex photoconductivity of the sample with fixed Fermi energy as a function of frequency for an excitation fluence
corresponding to $N_{\rm exc} = 8\cdot10^{12}$ carriers/cm$^2$,
together with the model result with a carrier heating efficiency
of $\eta$ = 0.5. This shows that in this regime other energy
relaxation channels contribute to the ultrafast energy relaxation.
The insets in panels \textbf{a} and \textbf{b} schematically show
the process of carrier heating through intraband scattering and
the corresponding carrier distributions for the ambient carrier
temperature $T_{\rm el}$ (blue) and the elevated carrier
temperature $T_{\rm el}'$ after photoexcitation and carrier
thermalization (red). The error bars in panels \textbf{a} and \textbf{b} represent the experimental (non-systematic) error bars (95 \% confidence interval).}
   \end{figure}

\clearpage

\begin{figure} [h!!!!!]
   \centering
   \includegraphics [scale=1.2]
   {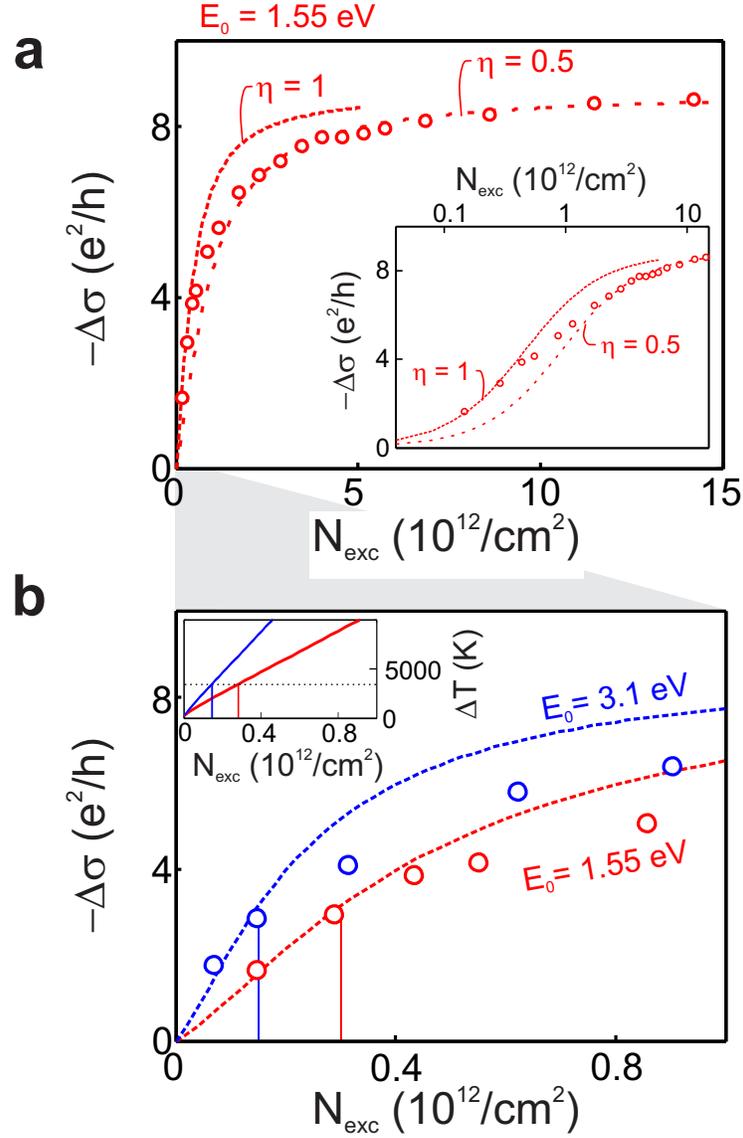}
      \caption{\textit{Carrier heating efficiency - dependence on fluence.}
\textbf{a)} The peak photoconductivity of the sample with fixed Fermi energy for a large
range of excitation powers and an excitation wavelength of 800 nm
together with the model with $\eta$ = 1 (short dashed line) and a
heating efficiency of $\eta$ = 0.5 (long dashed line). The inset shows the same data on a logarithmic horizontal scale.
\textbf{b)}  The peak photoconductivity of the sample with fixed Fermi energy as a function of excitation power ($N_{\rm exc}$) for
excitation with 800 nm light (corresponding to an energy of the
primary excited carrier of $E_0/2$ = 0.75 eV) and 400 nm light
($E_0/2$ = 1.5 eV). The dashed lines correspond to the model with
the same parameters (Fermi energy and scattering time proportionality constant) as in
panel $\textbf{a}$, using $\eta$ = 1. The model
describes the data well up to $N_{\rm exc} = 0.3 \cdot
10^{12}$ carriers/cm$^2$ for 800 nm excitation and up to $N_{\rm
exc} = 0.15 \cdot 10^{12}$ carriers/cm$^2$ for 400 nm excitation. The inset shows the carrier temperature that is reached after excitation and thermalization by carrier-carrier scattering, reaching $\sim$4000 K before heating becomes less efficient, for both excitation wavelengths. The symbol size of the data in panels \textbf{a} and \textbf{b} represents the error.}
   \end{figure}

\clearpage

\begin{figure} [h!!!!!]
   \centering
   \includegraphics [scale=1.2]
   {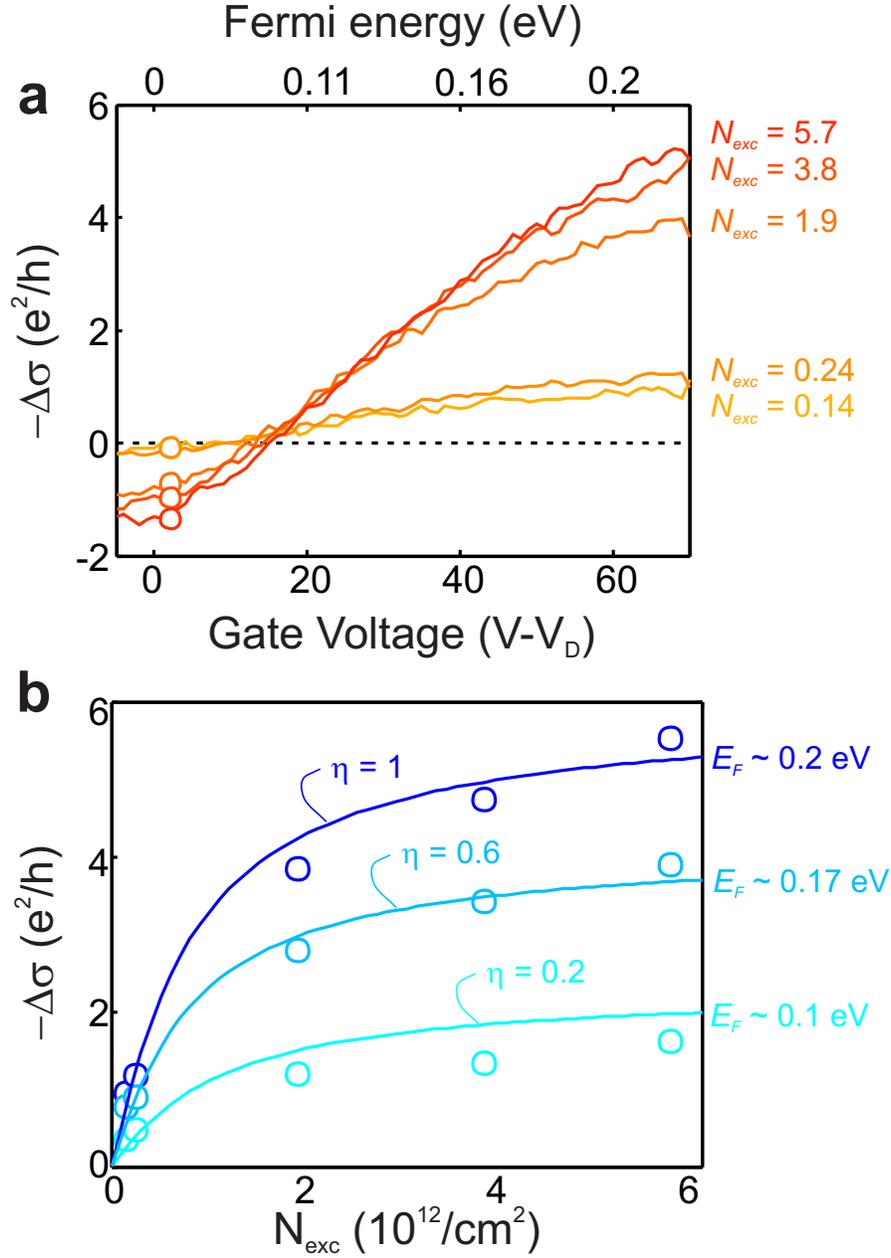}
      \caption{ \textit{Carrier heating efficiency -- dependence on Fermi energy.} \textbf{a)} The peak photoconductivity of the sample with controllable Fermi energy as a function of gate voltage for five different
excitation powers at excitation wavelength 1500 nm. The
corresponding excitation density $N_{\rm exc}$ is given in units
of $10^{12}$ absorbed photons/cm$^2$. At gate voltages away from
the Dirac point the photoconductivity is negative, while it changes
sign upon approaching the Dirac point, in agreement with very recent observations \cite{Shi2014,Frenzel2014}.
\textbf{b)} The peak photoconductivity as a function of excitation
power ($N_{\rm exc}$) for three gate voltages, corresponding to
$N_{\rm int}$ = 1, 2 and 3$\cdot10^{12}$ carriers/cm$^2$ (light to
dark blue). The solid lines show the results of the model
described in the text and the Methods section with a fixed heating
efficiency. For a decreasing number of intrinsic carriers (lower Fermi energy) the carrier heating efficiency decreases, in
agreement with the slowing down of the rise with decreasing Fermi energy. The symbol size of the data in panels \textbf{a} and \textbf{b} represents the error.}
   \end{figure}

\clearpage

\renewcommand{\thesection}{S\arabic{section}}
\renewcommand{\thefigure}{S\arabic{figure}}
\renewcommand{\theequation}{S\arabic{equation}}
\setcounter{figure}{0} \setcounter{section}{0}
\setcounter{equation}{0}

\section*{SUPPLEMENTARY MATERIAL}

\subsection{Characterization of the sample with controllable Fermi energy}

In order to translate the applied gate voltage $V_g$ into the
Fermi energy of the graphene sheet, we determine the capacitive
coupling  of the weakly doped silicon backgate to the graphene
sheet. For this, we use a sample with the same substrate as the
sample used for the $E_F$-dependent measurements, however with
graphene shaped as a Hall bar (see bottom right inset of Fig.\
S2a). We apply a current of $I = 1 \mu$A in the $x$-direction and
measure the Hall voltage $V_H$ between two contacts in the
$y$-direction. We measure the Hall voltage as a function of
backgate voltage for a $z$-directed magnetic field of both +0.6 T
and -0.6 T, and use the difference between these two scans to
account for offsets (from device asymmetry, for example). From the
Hall voltage we extract the carrier density $vs.$ backgate voltage
(see Fig.\ S1a) using

\be N_{\rm int} = {{I \cdot B}\over{e \cdot V_H}}, \ee

with $e$ the elementary charge. This allows us to extract the
capacitance of the backgate, where we find $C = 6.3 \cdot 10^{-5}$
F/m$^2$, which is somewhat smaller than the theoretical value for
300 nm oxide: $C_{\rm th} = 11.6 \cdot 10^{-5}$ F/m$^2$. The
reason for this could be the low density of carriers in the
silicon or the large size of the graphene flake. We use the
obtained capacitive coupling to obtain the Fermi level
corresponding to the applied gate voltages of the sample. It also allows us to obtain the device mobility, contact resistance and
 width of the neutrality point through the device resistance as a function of
 gate voltage. We find a contact resistance of 3.3 k$\Omega$, a neutrality point
 width of $\Delta$ = 58 meV, and a (lower bound) of the mobility of $\sim$300 cm$^2$/Vs
 (see Fig.\ S1b). This is in reasonable agreement with the $\sim$700 cm$^2$/Vs
 found from the photoconductivity fits.
 \\

\begin{figure} [h!!!!!]
   \centering
   \includegraphics [scale=0.7]
   {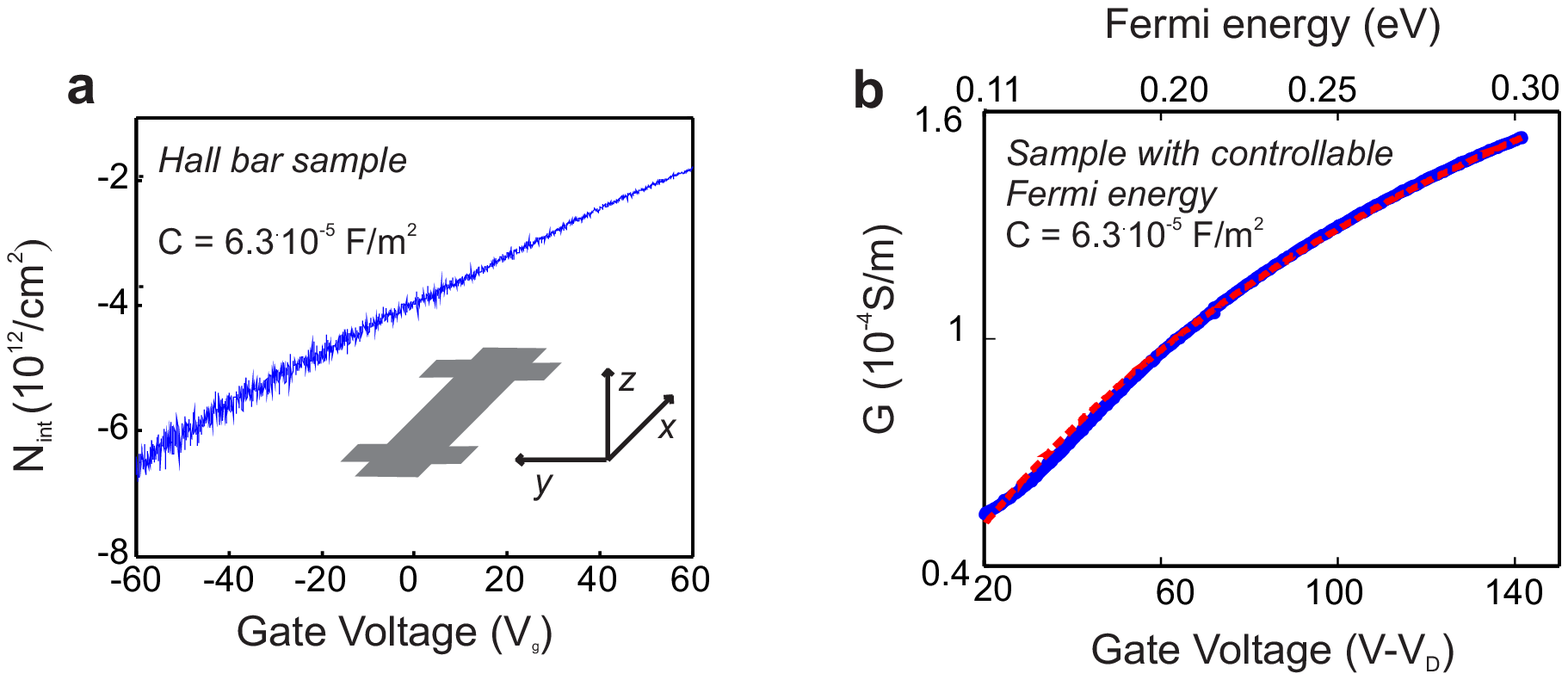}
      \caption{\textbf{a)} The number of intrinsic carriers as a function of gate voltage,
      extracted from Hall measurements on a similar device as the one used in the Fermi-level
      dependent measurements, equipped with additional contacts in the Hall geometry.
      \textbf{b)} The device resistance as a function of ($V_g$ - $V_D$) together with
       a fit using the determined capacitance, which yields the device mobility,
       contact resistance and neutrality region width. }
   \end{figure}

\subsection{Characterization of the sample with fixed Fermi energy}

We characterize the sample with fixed Fermi energy
using Raman spectroscopy  with pump wavelenth 532 nm (see an
exemplary Raman trace in Fig.\ S2a). The width of the 2D peak of
$\sim$33 cm$^{-1}$ shows that the graphene, grown by chemical
vapor deposition, is predominantly monolayer. We also extract the
Fermi energy, by analyzing the G peak location and the ratio
between the 2D and the G peak ($\sim$1585 cm$^{-1}$) and $>$2,
respectively, for the trace shown in Fig.\ S2). The Raman spectra
of more than 100 traces on different locations on the sample yield
an average G peak position of $\sim$ 1585 cm$^{-1}$, which
indicates a Fermi energy of $<$0.15 eV [S1, S2]. To get the most
realistic estimate for the Fermi energy that corresponds to our
optical pump - THz probe experiment, the Raman spectra are taken
under very similar environmental conditions (nitrogen flushing).
\\

\begin{figure} [h!!!!!]
   \centering
   \includegraphics [scale=0.7]
   {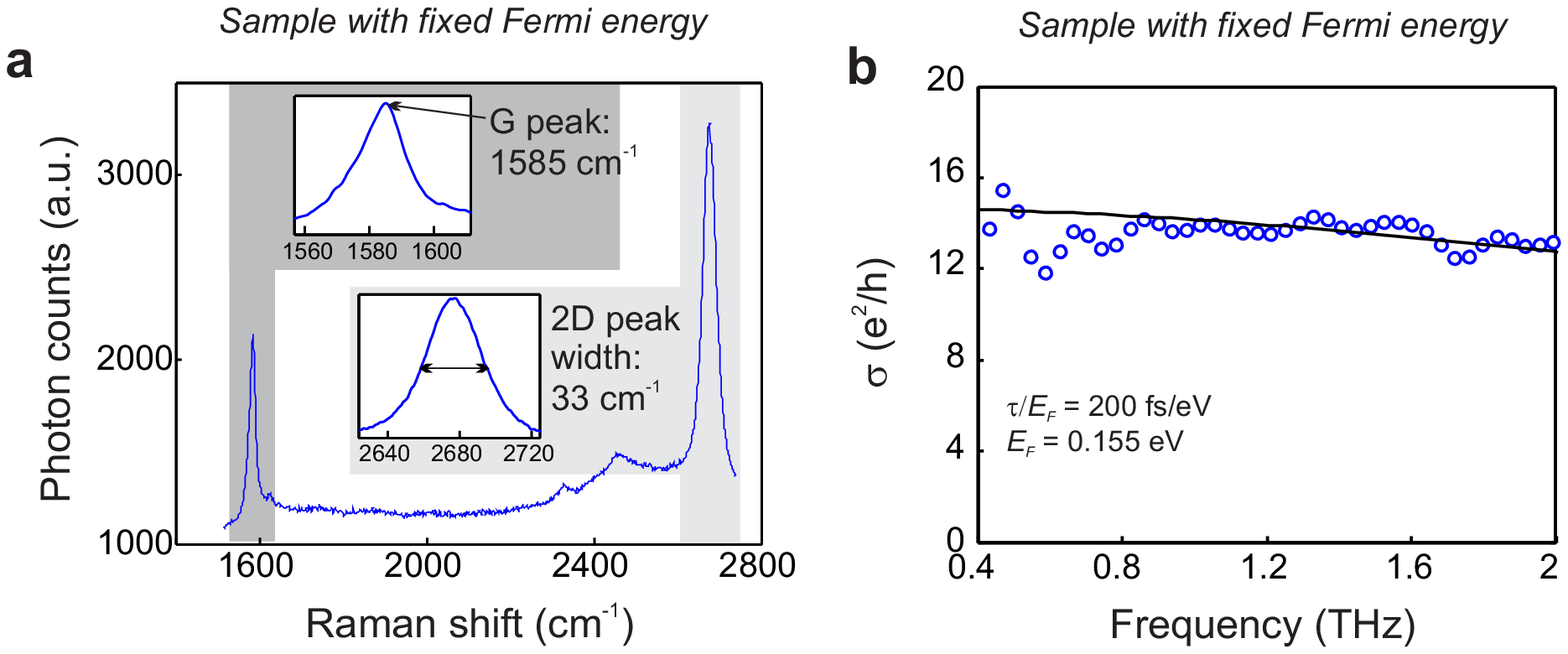}
      \caption{\textbf{a)} Raman spectrum of the sample with fixed Fermi energy.
      The insets show the G peak and the 2D peak. \textbf{b)} The steady state conductivity
      spectrum of the sample with fixed Fermi energy} (without photoexcitation), together with a Drude conductivity fit (black solid line).
   \end{figure}

We further characterize the sample using THz conductivity
measurements, where we  alternatively measure the substrate with
and without graphene in the THz focus. This allows us to obtain
the steady state conductivity (without photoexcitation). The
results are shown in Fig.\ S2b together with a conductivity fit as
explained in the Methods section. We obtain good agreement between
data and model for a Fermi level of $E_F =$ 0.155 eV and a
scattering time proportionality constant of 200 fs/eV, corresponding to a mobility of $\sim2300$ cm$^2$/Vs. We use this scattering time
proportionality constant and a slightly lower Fermi energy ($E_F = 0.11$ eV) to
compare all optical pump - terahertz probe data measured on this sample to the carrier heating
model. The slightly lower Fermi energy follows from the photoconductivity at high fluence in Fig.\ 4a (given the scattering time proportionality constant of 200 fs/eV), and could be due to spatial
variation of the Fermi energy, photo-cleaning during the course of
the experiment, or due to a modified humidity during the
experiment.
\\

\subsection{Positive photoconductivity at the Dirac point and the Silicon contribution}

\begin{figure} [h!!!!!]
   \centering
   \includegraphics [scale=0.7]
   {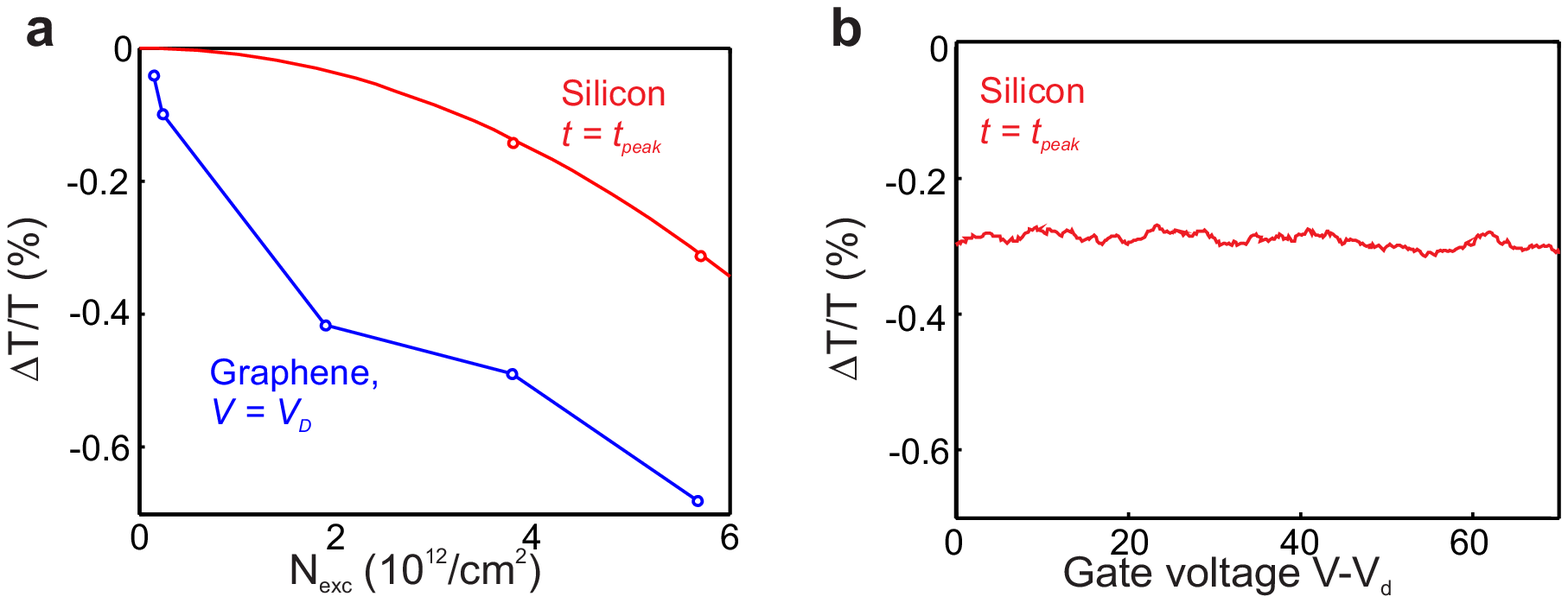}
      \caption{Pump-probe signal recorded at the pump-probe delay
       corresponding to the graphene signal peak, measured at the Dirac point
(blue) with negative change in transmission (positive
conductivity) as a function of absorbed fluence (in the graphene sheet). The same
measurement on the substrate without graphene at the same pump-probe delay gives a quadratic
signal due to two photon absorption (red). }
   \end{figure}

We use an excitation wavelength of 1500 nm for our pump-probe
measurements on the gated graphene sample to avoid exciting
electron-hole pairs in the silicon, which would obscure our
signal. However, there is a small contribution of two-photon
absorption that leads to THz photoconductivity in the silicon, in
addition to the photoconductivity of the graphene sheet. We find
that for the lowest fluences the silicon signal is negligible.
However due to the quadratic increase with fluence, the silicon
signal is not negligible  at the highest fluences, in cases where
the Fermi energy, and therefore the graphene signal, is very low.
In Fig.\ S3a we show the pump probe signal that originates from
the substrate without graphene and the pump probe signal for the
substrate with graphene at very low Fermi energy. It is noteworthy
to point out that the silicon contribution to the combined
graphene-silicon signal is insignificant at gate voltages far away
from the Dirac point and that it does not change with gate voltage
(see Fig.\ S3b). Only around the Dirac point, and at the highest
fluences, does part of the positive photoconductivity come from
the graphene and part from the silicon. Importantly, at low
fluence, the positive photoconductivity that we observe stems
completely from the graphene, showing the capability of tuning the
Fermi energy close enough to the Dirac point to change the sign of
the graphene THz photoconductivity. And even at the highest
employed fluence, the signal from graphene at the Dirac point is
still larger than the signal from the silicon substrate. We
corrected the fluence dependent results presented in Fig.\ 5 of
the main paper for this small substrate contribution.
\\

\subsection*{Supplementary References}

\noindent [S1] H. Yan et al., Infrared spectroscopy of wafer-scale graphene. \textit{ACS Nano} \textbf{5}, 9854--9860 (2011)

\noindent [S2] A. Das et al., Monitoring dopants by Raman scattering in an electrochemically top-gated graphene transistor. \textit{Nature Nanotech.}
\textbf{3}, 210--215 (2008)

\end{document}